\documentclass[aps,prd,reprint,twocolumn,preprintnumbers,superscriptaddress,nofootinbib]{revtex4-1}

\usepackage{amsmath}
\usepackage{graphicx}
\usepackage{float}
\usepackage[colorlinks=True, citecolor=blue, urlcolor=blue, linkcolor=blue]{hyperref}

\usepackage{amssymb}
\usepackage[normalem]{ulem}

\begin{document}

\title{Diffusion model approach to simulating electron-proton scattering events}

\author{Peter Devlin}
\email{devlin@jlab.org}
\affiliation{Thomas Jefferson National Accelerator Facility, Newport News, VA 23606, USA}

\author{Jian-Wei Qiu}
\email{jqiu@jlab.org}
\affiliation{Thomas Jefferson National Accelerator Facility, Newport News, VA 23606, USA}

\author{Felix Ringer}
\email{fmringer@jlab.org}
\affiliation{Thomas Jefferson National Accelerator Facility, Newport News, VA 23606, USA}
\affiliation{Department of Physics, Old Dominion University, Norfolk, VA 23529, USA}

\author{Nobuo Sato}
\email{nsato@jlab.org}
\affiliation{Thomas Jefferson National Accelerator Facility, Newport News, VA 23606, USA}

\preprint{JLAB-THY-23-3945}

\date{\today}

\begin{abstract}
Generative AI is a fast-growing area of research offering various avenues for exploration in high-energy nuclear physics. In this work, we explore the use of generative models for simulating electron-proton collisions relevant to experiments like CEBAF and the future Electron-Ion Collider (EIC). These experiments play a critical role in advancing our understanding of nucleons and nuclei in terms of quark and gluon degrees of freedom. The use of generative models for simulating collider events faces several challenges such as the sparsity of the data, the presence of global or event-wide constraints, and steeply falling particle distributions. In this work, we focus on the implementation of diffusion models for the simulation of electron-proton scattering events at EIC energies. Our results demonstrate that diffusion models can reproduce relevant observables such as momentum distributions and correlations of particles, momentum sum rules, and the leading electron kinematics, all of which are of particular interest in electron-proton collisions. Although the sampling process is relatively slow compared to other machine learning architectures, we find diffusion models can generate high-quality samples. We foresee various applications of our work including inference for nuclear structure, interpretable generative machine learning, and searches of physics beyond the Standard Model.
\end{abstract}

\maketitle
%\tableofcontents

%

\section{Introduction}

\begin{figure}[t]
    \centering   
    \includegraphics[width=0.45\textwidth]{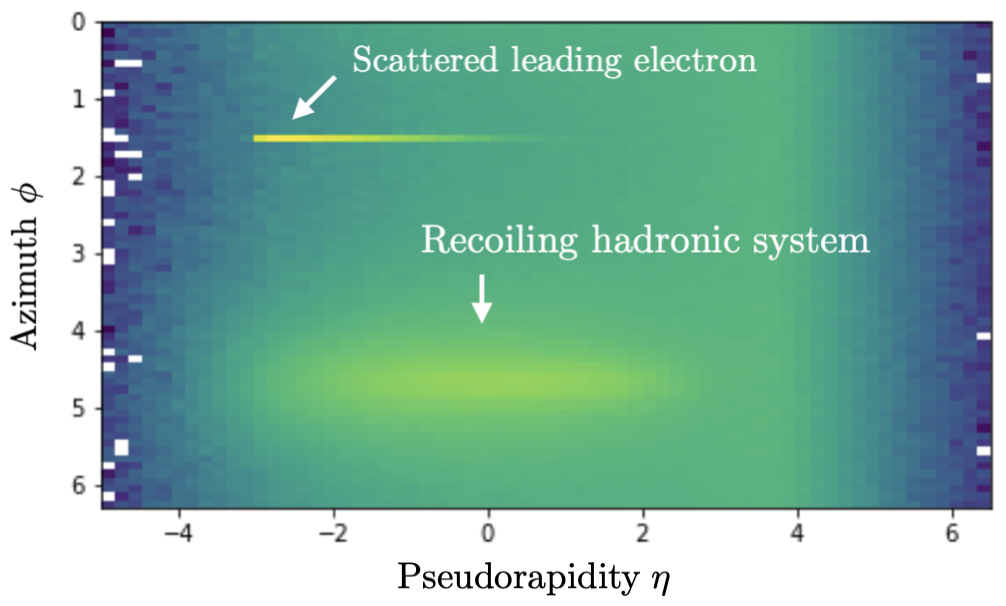}
\caption{Momentum distribution of the particles in the $\eta$-$\phi$ plane created in electron-proton scattering events in the laboratory frame at $\sqrt{s}=105$~GeV. The events that have been generated with \textsc{Pythia8}~\cite{Sjostrand:2014zea} are rotated such that the scattered electron is in the same azimuthal direction for all events as indicated in the figure.~\label{fig:ep_events}}
\end{figure}

High-energy particle and nuclear collider experiments along with theoretical progress in the past decades have allowed for an increasingly sophisticated understanding of the quark and gluon dynamics at subatomic scales. Electron-proton scattering experiments including HERA at DESY, CEBAF at JLab, and the future Electron-Ion Collider (EIC) at BNL~\cite{AbdulKhalek:2021gbh} and LHeC~\cite{LHeC:2020van}/FCC-eh at CERN~\cite{FCC:2018byv} play a critical role in advancing our understanding of the structure of hadrons, probing cold nuclear matter effects, and searching for physics beyond the Standard Model. In particular, the measurement of the scattered leading electron provides a clean electromagnetic probe of the inner structure of hadrons and nuclei. The experimental data has been analyzed within the framework of QCD factorization to extract the three-dimensional structure of hadrons in terms of quantum correlation functions, such as parton distribution functions (PDFs). In addition, collider studies related to the emergence of hadrons and the associated neutralization of color have remained at the forefront of collider experiments. See Fig.~\ref{fig:ep_events} for an illustration of the average distribution of particles in high-energy electron-proton collisions, which will be discussed in more detail below.

\begin{figure*}[t]
    \centering   
    \includegraphics[width=0.95\textwidth]{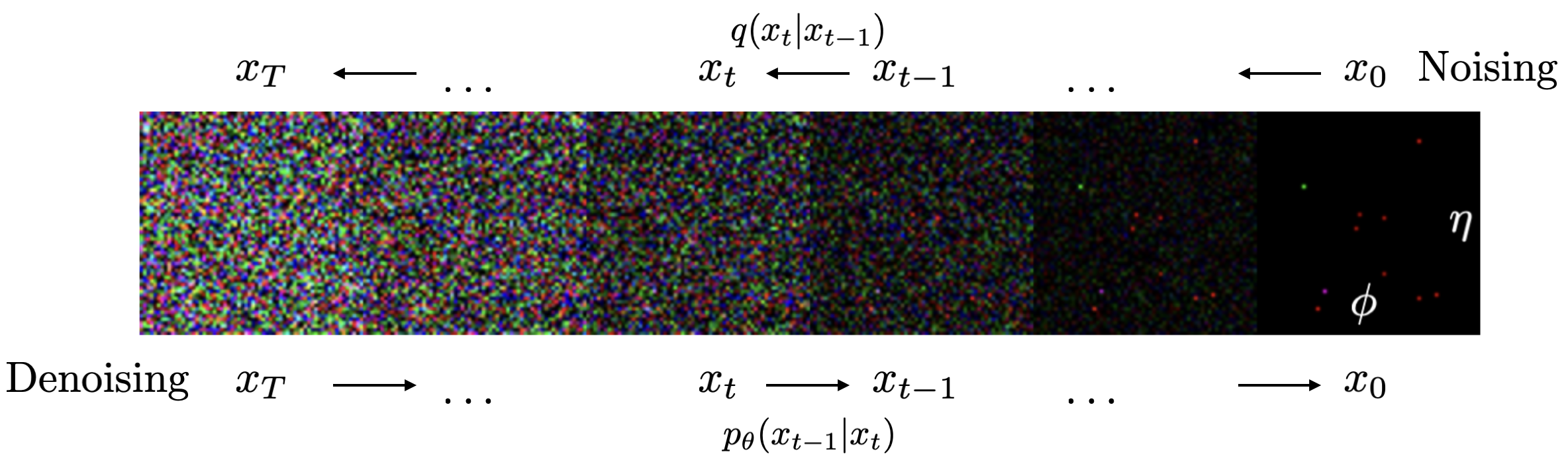}
\caption{Sequence of images illustrating the noising/denoising process of a diffusion model trained on \textsc{Pythia8} simulations of electron-proton scattering events. Pixels colored in black are empty, and the three RGB color channels correspond to charged pions $\pi^+$, electrons $e^-$, and kaons $K^+$, respectively.~\label{fig:diffusion}}
\end{figure*}

The rapid development of AI and machine learning in recent years has led to a wide range of applications in particle and nuclear physics~\cite{Carleo:2019ptp,Boehnlein:2021eym}. Examples include the simulation of lattice gauge configurations~\cite{Albergo:2019eim,Nicoli:2020njz,Lawrence:2021izu,Abbott:2023thq,Demirtas_2024}, the classification of jets~\cite{deOliveira:2015xxd,Kasieczka:2017nvn,Datta:2017rhs,Komiske:2018cqr,Heimel:2018mkt,Dreyer:2021hhr,Cai:2021hnn,Lee:2022kdn,Araz:2022haf,Athanasakos:2023fhq}, the simulation of collider events~\cite{deOliveira:2017pjk,Butter:2019cae,Gao:2020zvv,Danziger:2021eeg,Butter:2022rso}, the unfolding of detector effects~\cite{Bellagente:2019uyp,Andreassen:2019cjw,Alanazi:2020jod,Alghamdi:2023emm,Huang:2023kgs}, data analyses with machine learning-improved Bayesian posterior sampling~\cite{He:2018gks,https://doi.org/10.48550/arxiv.2107.08001,JETSCAPE:2020mzn,Gabrie:2021tlu,Hunt-Smith:2023ccp,Yamauchi:2023xrz}, regression tasks~\cite{NNPDF:2017mvq,Grigsby:2020auv,DelDebbio:2020rgv,Arratia:2021tsq,Diefenthaler:2021rdj,Fanelli:2023lmp}, and searches of physics beyond the Standard Model~\cite{Nachman:2020lpy,Andreassen:2020nkr,Finke:2021sdf,Fraser:2021lxm,Araz:2022zxk,Morandini:2023pwj}. See Ref.~\cite{Feickert:2021ajf} for a broad overview.
Several of these applications rely on generative models that can learn the structure or latent space of a data set and generate new samples. Various types of generative models have been developed including  Variational Autoencoders~\cite{2013arXiv1312.6114K}, Autoregressive Models~\cite{DBLP:journals/corr/abs-1906-00446}, Generative Adversarial Networks (GANs)~\cite{Goodfellow:2014}, flow based models~\cite{0835426dfb054a17b6dd75c4f5f98961} and diffusion models~\cite{DBLP:journals/corr/Sohl-DicksteinW15,DBLP:journals/corr/abs-2006-11239}. The different types of generative models each have their own advantages and disadvantages. The choice of generative models for a particular application depends for example on the computational cost of training and sampling from the model, the quality of the generated samples, the scalability, and the stability of the training procedure, etc. In this work, we implement a diffusion model, which can generate samples from a data distribution by learning to reverse a stepwise noising or diffusion process, see Fig.~\ref{fig:diffusion}. While sampling from diffusion models is generally relatively slow compared to other architectures, they have been shown to generate high-quality samples and allow for a scalable and stable training procedure. For example, in Ref.~\cite{DBLP:journals/corr/abs-2105-05233} it was found that diffusion models outperform GANs in image synthesis. The ability of diffusion models to generate high-quality samples is essential for the applications we foresee in the context of high-energy collider physics. In addition, Ref.~\cite{DBLP:journals/corr/abs-2102-09672} reported that diffusion models may cover a larger portion of the target distribution as GANs. 

The development of generative models for simulating collider events or jets, collimated sprays of particles, was first initiated with GANs in Refs.~\cite{deOliveira:2017pjk,Paganini:2017dwg,Alanazi:2020klf}. Since then, different architectures such as normalizing flows~\cite{Krause:2021ilc} have been explored as well as different data representations such as point clouds instead of images have been considered~\cite{Leigh:2023toe,Butter:2023fov,Acosta:2023zik,Leigh:2023zle,Buhmann:2023pmh}. In addition, efforts have been made to increase the interpretability of generative models~\cite{Lai:2020byl,Bieringer:2020tnw}. One of the challenges for generative models is the sparsity of collider events, which is distinct from typical tasks encountered in computer vision. In addition, event-wide constraints such as momentum conservation and steeply-falling momentum distributions of particles add to the complexity of the problem. Recently, diffusion and score-based generative models~\cite{DBLP:journals/corr/abs-1907-05600} have been developed for simulations of calorimeter showers~\cite{Mikuni:2022xry,Mikuni:2023dvk,Amram:2023onf,Buhmann:2023zgc,Imani:2023blb} and jets using point clouds~\cite{Leigh:2023toe,Butter:2023fov,Acosta:2023zik,Mikuni:2023tqg,Leigh:2023zle}. In this work, we will simulate full collider events with diffusion models focusing in particular on the unique characteristics of electron-proton (and similarly electron-nucleus) collisions. For example, the kinematics of the leading electron play a critical role since it is used to determine both the photon virtuality $Q^2$ and the scaling variable Bjorken $x$ of the event. In addition, particle spectra span up to 6 orders in magnitude and, depending on the type of particle, they peak in different regions of phase space. We address this challenge by using a suitable preprocessing step before training the model on the data. In addition, we explore different loss functions and optimization procedures of diffusion models. Our findings demonstrate that diffusion models are able to generate high-quality samples indicating their potential for various future applications in high-energy nuclear and particle physics.

We foresee various applications of generative models for collider events. For example, generative models are closely related to the development of parton showers and Monte Carlo event generators. While the perturbative part is increasingly well understood from first principles in QCD~\cite{Dasgupta:2020fwr,Forshaw:2020wrq,Herren:2022jej,Neill:2021std}, other components of Monte Carlo event generators can be simulated with generative models, see for examples the approach developed in Ref.~\cite{Lai:2020byl}. Nuclear physics applications include the modification of the shower due to hot or cold nuclear matter. Moreover, anomaly detection techniques based on generative models have been developed for searches of physics beyond the Standard Model. The identification of anomalous signals requires an accurate modeling of the background distribution. Generative modeling also finds applications in hadron structure studies, which are a prominent subject of research within the Jefferson Lab 12 GeV program and the future EIC. The increasing sophistication required to extract parton-level information including transverse momentum distributions (TMDs) and generalized parton distributions (GPDs), necessitates the analysis of multi-dimensional phase space distributions from semi-inclusive and exclusive observables. In this context, generative modeling can be used as a generator of partonic structures for QCD global analysis, a phase space generator for particle reactions, or emulators for detector simulation \cite{Alghamdi:2023emm, Alanazi:2020jod}. It is also opening new avenues to integrate theory and experiment within a unified event-level analysis.

The remainder of this paper is organized as follows. In section~\ref{sec:diffusion}, we provide a review of diffusion models. In section~\ref{sec:data}, we discuss the generation of electron-proton scattering events with \textsc{Pythia8}~\cite{Sjostrand:2014zea}, which is subsequently used as the training data, as well as the data representation. In section~\ref{sec:implementation}, we provide details of our implementation and the training procedure. In section~\ref{sec:numerics}, we present numerical results comparing the diffusion model to \textsc{Pythia8} using various metrics relevant to simulating collider events. Lastly, we conclude and present an outlook in section~\ref{sec:conclusions}.

\section{Diffusion models~\label{sec:diffusion}} 

Diffusion models are a class of generative machine learning models that can learn the underlying distribution of a given data set. The training procedure of diffusion models consists of two components -- a noising, and a denoising process, see Fig.~\ref{fig:diffusion}. Starting with pixelated images of the training data set, noise is incrementally added to the image until it is ultimately transformed into pure noise. Subsequently, the inverse denoising process can be learned by a suitably chosen machine learning architecture. Due to the stepwise nature of the diffusion process, the results of the entire chain can be included in the loss function allowing for a scalable training process. After the training procedure is finished, we can generate new samples of the target data set by passing noise through the trained machine learning architecture.

Before starting the diffusion process, the pixelated images are treated as data vectors, which we label as $x_0$. The corresponding probability distribution of the data is $x_0\sim q_0(x_0)$. Analogously, we denote the data vector at time step $t$ of the diffusion process as $x_t$, with $t\in[0,T]$, and the corresponding probability distribution is $x_t\sim q_t(x_t)$. The stepwise forward diffusion or noising process is given by
\begin{equation}\label{eq:markov}      
    q(x_1,\ldots,x_T|x_0)=\prod_{t=1}^Tq(x_t|x_{t-1})\,.
\end{equation}
The probability distribution for a given timestep of the noising process $x_{t-1}\to x_t$, is given by
\begin{equation}\label{eq:noising}
    q\left(x_t | x_{t-1}\right)=\mathcal{N}(x_t ; \sqrt{1-\beta_t} x_{t-1}, \beta_t \mathbf{I})\,.
\end{equation}
Here ${\cal N}$ is a multi-variate Gaussian distribution with a diagonal covariance matrix. The values of $\beta_t$ are chosen according to a predefined variance schedule $\{\beta_t\in (0,1)\}_{t=1}^T$. We are thus adding a certain amount of Gaussian noise at each time step leading to a sequence of increasingly noisy samples $x_0,\ldots,x_T$, where the variance schedule and the time steps are chosen such that $x_T$ is eventually an isotropic Gaussian distribution $q(x_T)={\cal N}(x_T;0,\mathbf{I})$. Note that Eqs.~(\ref{eq:markov}) and (\ref{eq:noising}) describe a Markovian process since the probability distribution at time step $t$ only depends on the current sample at time $t-1$.

Next, we consider the reverse diffusion or denoising process. We need to train a suitable machine-learning model to approximate the probability distribution of the inverse process $q(x_{t-1}|x_t)$. The diffusion process is stochastic, which does not allow for the use of backpropagation techniques to obtain the gradient. Instead, the reparametrization trick is used to make the problem tractable and learn the parameters of a Gaussian distribution for which backpropagation can be used~\cite{2013arXiv1312.6114K}. The denoising process $x_{t}\to x_{t-1}$ proceeds again in $T$ time steps where the following Gaussians are sampled from
\begin{equation}\label{eq:denoising}
    p_\theta\left(x_{t-1} | x_t\right)=\mathcal{N}\left(x_{t-1} ; \mu_\theta\left(x_t, t\right), \Sigma_\theta\left(x_t, t\right)\right) \,.
\end{equation}
The mean $\mu_\theta$ and covariance $\Sigma_\theta$ are learned by the model, where $\theta$ denotes the trainable parameters. Typically a U-shaped convolutional neural network (U-Net) is used as a model to learn the mean and variance at each time step $t$~\cite{DBLP:journals/corr/RonnebergerFB15}. 

The model parameters are obtained by minimizing a loss function during the training procedure. Different options have been explored in the literature. We start by considering the Variational Lower Bound (VLB), which can be written as follows
\begin{align}\label{eq:VLB}
L_{\mathrm{VLB}} & =L_0+L_1+\ldots+L_{T-1}+L_T \\
L_0 & =-\log p_\theta\left(x_0 |x_1\right) \,,\\
L_{t-1} & =D_{\rm K L}\left(q\left(x_{t-1} | x_t, x_0\right) \|\, p_\theta\left(x_{t-1} | x_t\right)\right) \,, \\
L_T & =D_{\rm K L}\left(q\left(x_T | x_0\right) \| \, p\left(x_T\right)\right)\,.
\end{align}
Here $L_{t-1}$ is used for all terms in Eq.~(\ref{eq:VLB}) except for $t=0,T$. Except for $L_0$, closed-form expressions can be found for all KL divergences since each term involves two Gaussian distributions. In Ref.~\cite{DBLP:journals/corr/abs-2006-11239}, it was found empirically that a simplified objective function can improve the sample quality. Instead of using a neural network to predict $\mu_\theta$ and $\Sigma_\theta$, the network is used to predict $x_0$ and the noise $\epsilon_\theta$ at each time step, which can be related to $\mu_\theta$, see Ref.~\cite{DBLP:journals/corr/abs-2006-11239} for more details. The simplified mean squared error objective function, is given by
\begin{equation}
    L_t^{\text {simple }}=E_{t, x_0, \epsilon}[\left\|\epsilon-\epsilon_\theta\left(x_t, t\right)\right\|^2]\,,
\end{equation}
where $\epsilon$ represents the noise of the forward diffusion process. Since this simplified objective function is only sensitive to $\mu_\theta$ but not $\Sigma_\theta$, a hybrid loss function was introduced in Ref.~\cite{DBLP:journals/corr/abs-2102-09672}
\begin{equation}\label{eq:hybrid}
    L_{\rm hybrid} = L_{\rm simple} + \lambda L_{\rm VLB} \,.
\end{equation}
Here, $\lambda$ is a hyperparameter that determines the relative importance of the two objective functions. Typically, $\lambda$ is chosen to be relatively small such that $\mu_\theta$ is primarily determined by $L_{\rm simple}$ and  $\Sigma_\theta$ is related to $L_{\rm VLB}$. In section~\ref{sec:implementation}, we will provide more details of the setup used in this work. 

After the training procedure, we can obtain new samples $q(x_0)$ from the target distribution by sampling $x_T\sim {\cal N}(0,\mathbf{I})$ and running the reverse process of the diffusion model. The Markovian noising and denoising processes described in Eqs.~(\ref{eq:noising}), (\ref{eq:denoising}) are analogous to the diffusion process in non-equilibrium thermodynamics and in the continuous-time limit, a stochastic differential equation is obtained~\cite{song2021scorebased}. See also Ref.~\cite{weng2021diffusion} for a more detailed introduction to diffusion models.

\section{Training data set and data representation~\label{sec:data}} 

We generate the training data set, by simulating neutral-current electron-proton scattering events with \textsc{Pythia8}~\cite{Sjostrand:2014zea} using $\sqrt{s}=105$~GeV as a representative center-of-mass (CM) energy for the future EIC~\cite{AbdulKhalek:2021gbh}. Since the photoproduction region (low photon virtuality $Q^2$), and deep inelastic scattering (DIS) region (high $Q^2$) are sensitive to different physics, we choose to impose a lower cut of $Q^2>25$~GeV$^2$ to exclude the photoproduction events. Fig.~\ref{fig:ep_events} shows a 2D histogram of the momentum distribution of particles in electron-proton scattering events in the laboratory frame. We highlight in particular the kinematic region of the scattered electron. Note that the electron pseudorapidity does not extend to very low values due to the $Q^2$ cut. In addition, we indicate the recoiling hadronic system or jet, which is produced in the opposite azimuthal direction as the scattered electron. For each generated particle $i$ in the event, we record the transverse momentum $p_{Ti}$ relative to the beam axis, the pseudorapidity $\eta_i=-\ln\tan\theta/2$ with the polar angle $\theta_i$ with respect to the direction of colliding electron, the azimuthal angle $\phi_i$ and the particle identification (PID$_i$). We impose a cut on the pseudorapidity of $|\eta_i|<10$, which captures the vast majority of particles produced by \textsc{Pythia8}, see Fig.~\ref{fig:ep_events}. We do not apply a lower cut on the transverse momentum $p_{Ti}$ of the particles. Instead of directly incorporating the transverse momentum of each particle as a feature of the training data set, we choose to work with a rescaled variable. A natural choice for the rescaled momentum variable is
\begin{equation}\label{eq:sum}
    \tilde z_i=\frac{2M_{Ti}}{\sqrt{s}}\cosh y_i \,.
\end{equation}
Here $y_i$ is the rapidity and $M_{Ti}^2=p_{Ti}^2+m_i^2$ is the transverse mass, where $m_i$ is the hadron mass. This variable is of great interest for simulating full collider events since it satisfies the following event-wide momentum sum rule
\begin{equation}\label{eq:sumz}
    \Sigma \equiv \sum_{i\in{\rm event}}\tilde z_i=2 \,,
\end{equation}
where we sum over all particles in a given event. This provides an important global constraint for simulating full collider events. However, in practice, this requires having access to all particle species and fully hermetic detectors. In general, the rapidity coverage of detectors is limited and, in this work, we also limit the diffusion model setup to simulating only three particle species. As a result, Eq.~(\ref{eq:sumz}) is not exactly satisfied. Instead, Eq.~(\ref{eq:sumz}) provides an upper bound on the sum over all $\tilde z_i$ values in each event. In the limit of massless particles, $\tilde z_i$ reduces to
\begin{figure}[t]
    \centering   
    \includegraphics[width=.47\textwidth]{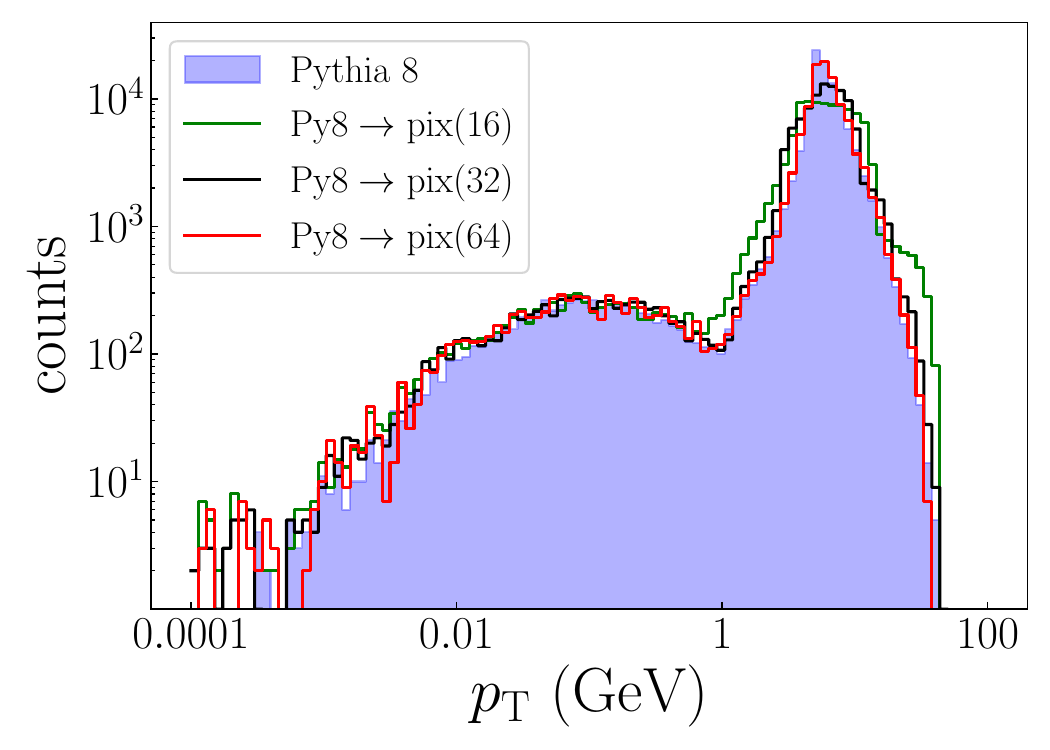}
\caption{Momentum distribution of electrons $e^-$ in electron-proton scattering events for three different levels of pixelation compared to the original \textsc{Pythia8} result.~\label{fig:pixels}}
\end{figure}
\begin{equation}\label{eq:zmassless}
    z_i=\frac{2p_{Ti}}{\sqrt{s}}\cosh \eta_i \,,
\end{equation}
where $\eta$ is the pseudorapidity, as introduced above. This variable is frequently used in the perturbative QCD literature, see for example Refs.~~\cite{Kaufmann:2015hma,Neill:2021std}. Therefore, we choose $z_i$ as the rescaled momentum variable for this work. Eq.~(\ref{eq:sumz}) also provides an upper bound for the sum over all $z_i$ in each event. In summary, we record the variables $(z_i,\eta_i,\phi_i,{\rm PID}_i)$ for each particle in the event. When training the diffusion model, as described below, we limit ourselves to only three ``color'' channels for which we choose: pions $\pi^+$, kaons $K^+$, and electrons $e^-$ from the full \textsc{Pythia8} generated event. In particular, the (leading) scattered electron in the event plays an important role in electron-proton scattering and we will study its kinematic distributions in detail below. Since we are primarily interested in a proof-of-concept study, we limit ourselves to modeling only a few representative particle species. In particular, the leading electron kinematics differ significantly from the rest of the hadronic system.

A convenient way to digitize collider events is to represent them as images, which is well suited for machine learning applications. We partition the cylindrical detector around the scattering vertex into a grid of uniformly sized rectangular pixels located at regular intervals in both rapidity and azimuth. We choose the pixel intensity to be the rescaled particle momentum $z_i$, and the rapidity $\eta_i$ and azimuthal angle $\phi_i$ index the location of each pixel on the cylinder. The particle type (PID$_i$) is stored as indices of the different image color channels similar to RGB color channels. Whenever multiple particles of the same type are in the same pixel, their $z_i$ values are added. 

In an actual experiment, the natural choice for the pixel intensity is $p_{Ti}$ since $z_i$ is a derived quantity given in terms of the measured value of $p_{Ti}$ and $\eta_i$. Therefore, combining multiple particles in a given pixel will occur at the level of $p_{Ti}$ instead of $z_i$. However, as we will discuss in section~\ref{sec:numerics} below, the reconstruction of physical observables in inclusive DIS scattering experiments is strongly affected by distortions induced by the $\eta_i$ pixelation. In part, this is due to the large pseudorapidity interval chosen for this work. In principle, this can be mitigated by increasing the number of pixels for the rapidity. Due to limited computing resources, we were not able to increase the number of pixels further in this work. In future work, this can be addressed by increasing the number of pixels or by changing the data representation. Here, we opt for using the $z_i$ as the pixel intensities, which reduces pixelation effects.

We quantify the discretization effect by considering as an example the inclusive momentum distribution of electrons $e^-$ in \textsc{Pythia8}-generated electron-proton scattering events, which is shown in Fig.~\ref{fig:pixels}. We observe a large-$p_{T}$ peak due to the leading scattered electron and a continuous spectrum at intermediate to small-$p_{T}$ values due to electrons generated during the shower. We compare the actual distribution with its discretized counterparts using three different choices for the number of pixels for the image: $16\times 16$, $32\times 32$, and $64\times 64$. We observe that the actual distribution is increasingly well reproduced as the number of pixels is increased. Throughout this work, we use $64\times 64$ pixels as our default choice. With larger computing resources this can be increased until the experimental resolution of the detector is reached.

As mentioned in the introduction, different than typical tasks in computer vision, images of collider events are very sparse, especially at the relatively low energies of the CEBAF experiment at JLab and the future EIC. For a CM energy of $\sqrt{s}=105$~GeV, we find that the average level of sparsity or the percentage of empty pixels is $99.95\pm 0.02\%$ (including all particle species) for $64\times 64$ images, which can be challenging for generative models. We address this problem by choosing a suitable data representation as discussed in the following. Besides the sparsity of the data, a significant challenge is the steeply falling distributions of particles. The inclusive momentum $z$-distributions peak close to the endpoints $z\to 0$ for $\pi^+, K^+$ since soft hadrons have a large production cross section in QCD. Instead, for electrons $e^-$ the distribution peaks in the region $z\to 1$, see Fig.~\ref{fig:pixels}. The large-$z$ peak of the electron/positron momentum distribution is a unique feature of electron-proton scattering events. Instead, in proton-proton collisions, all distributions peak at small-$z$ values. This feature appears due to the scattered leading electron, which plays a unique role in electron-proton collisions since it is used to determine the virtuality of the exchanged photon $Q^2$ and Bjorken $x$. Therefore, the accurate modeling of its kinematics plays a critical role. In order to take into account the logarithmic behavior of the data near both endpoints $z\to 0,1$ and to improve the training procedure, we rescale $z$ as follows
\begin{equation}\label{eq:rescale}
        z \to S(z)E(z) + (1-S(z))L(z) \,.
\end{equation}
Here $E(z)$ is an exponential function, $L(z)$ is a logarithmic function, and $S(z)$ is a sigmoid. The three functions are defined as 
\begin{align}
    E(z) & = a_1 e^{c_1 z}+b_1\,,\\
    L(z) & = a_2\ln(z+c_2)+b_2\,,\\
    S(z) & = \frac{1}{1+e^{-\alpha (z - \beta)}}\,,
\end{align}
where we have introduced additional parameters that will be discussed in the following. First, we require that the parameters are chosen such that Eq.~(\ref{eq:rescale}) is a bijective function allowing us to eventually recover to the original momentum distribution. Second, we choose the parameters such that the rescaling in Eq.~(\ref{eq:rescale}) matches the peak structures of the $z$ distributions near both endpoints. Note that we apply the same $z$ rescaling to all three channels. We choose the rescaling to be linear in the intermediate $z$ region while near the upper (lower) endpoint, the function approximates an exponential (logarithmic) function. This can be achieved by choosing two values $z_1<z_2\in  ]0,1[$ with $z_2$ being the value where the exponential $E(z)$ smoothly becomes a linear function, i.e.
\begin{equation}
    E(z_2) = z_2\,,\quad \frac{{\rm d}E(z_2)}{{\rm d}z}=1 \,.
\end{equation}
Similarly, we require the logarithmic function $L(z)$ to become linear at $z_1$. These conditions along with the need to construct a bijective function fixes or constrains several of the parameters in Eq.~(\ref{eq:rescale}). We then choose the sigmoid $S(z)$ to smoothly interpolate between the exponential and logarithmic functions. The remaining parameters are chosen such that the rescaled $z$ distribution is sufficiently smooth for the training of the diffusion model. We note that without the preprocessing of the training data described, the results of the diffusion model can be off by several orders of magnitude near the kinematic endpoints.

\begin{figure}[t]
    \centering   \includegraphics[width=.47\textwidth]{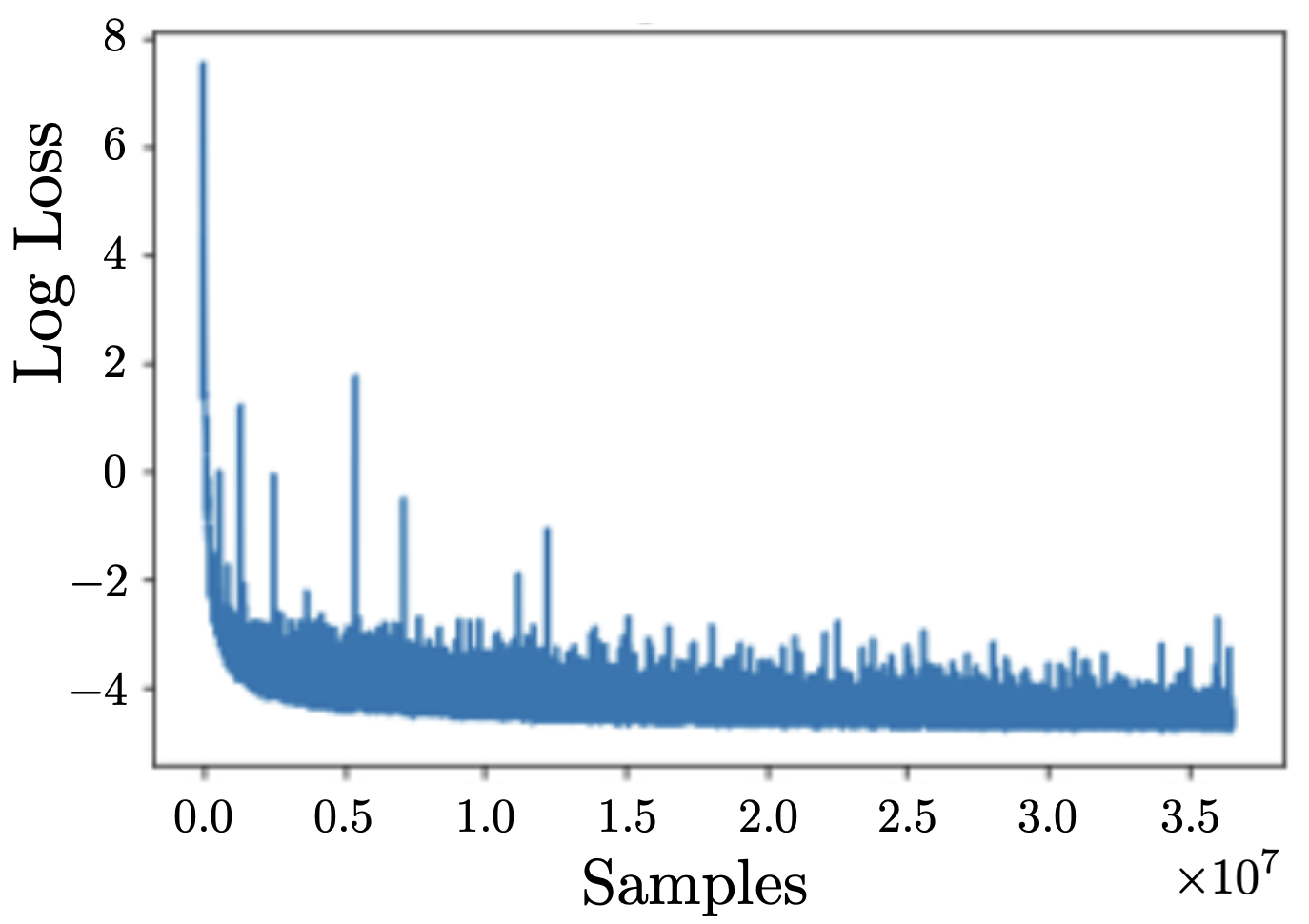}
    \caption{The log loss of the diffusion model training procedure as a function of the number of samples that the model is trained on.~\label{fig:loss}}
\end{figure}

\begin{figure*}[t]
    \centering   
    \includegraphics[width=\textwidth]{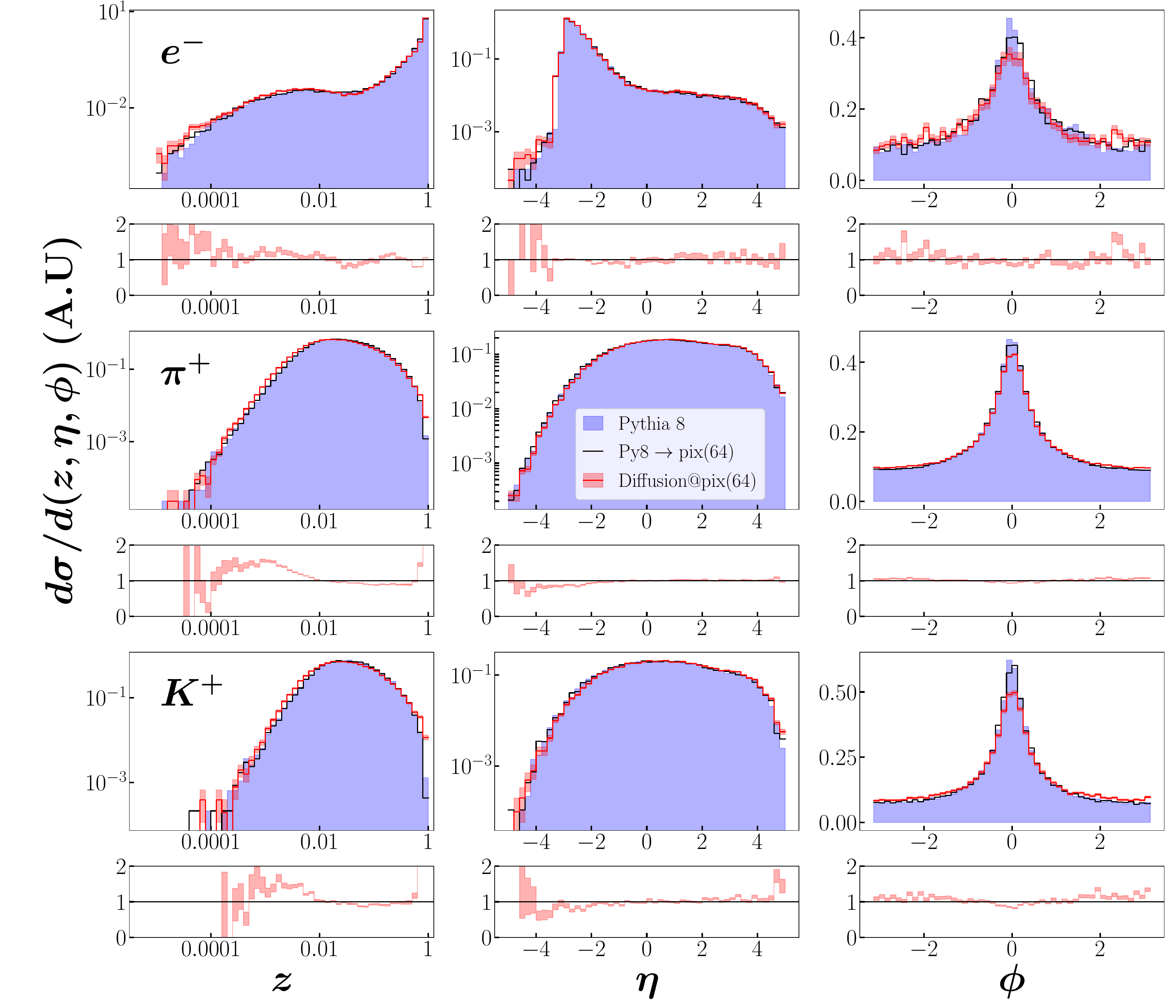}
    \caption{Differential cross sections in arbitrary units (A.U) of the rescaled momentum variable $z$, pseudorapidity $\eta$, and azimuthal angle $\phi$ (left to right). The results are shown separately for electrons $e^-$, charged pions $\pi^+$, and kaons $K^+$ (top to bottom). We show the diffusion model results (red) including statistical uncertainties and compare them to \textsc{Pythia8} with (black) and without (purple) pixelation effects. Under each panel, we include cross sections ratios between the diffusion model and \textsc{Pythia8} with pixelation.~\label{fig:momentum}} 
\end{figure*}

The diffusion model takes as input values of the momentum fraction in the range $[-1,1]$ (floating point numbers). Before pixelation and the rescaling in Eq.~(\ref{eq:rescale}), the range of the particle momentum fractions is in the range of $z\sim [10^{-6},1]$. We choose a suitable range of values $z'\in [z'_{\rm min},z'_{\rm max}]$ with $-1<z'_{\rm min}<z'_{\rm max}<1$ to which we map the original $z$ values. This includes the rescaling in Eq.~(\ref{eq:rescale}) as well as an additional linear transformation to match the targeted range. In practice, we find that $z'_{\rm min}=-0.76$ and $z'_{\rm max}=0.86$ work well for the purposes of this work. These values are chosen to allow for an upper and lower gap from the endpoints at $-1,1$. The lower gap allows us to train a diffusion model that can generate empty pixels. This is achieved by mapping empty pixel values in the training data set, i.e. initially at $z=0$, to the lower end of the allowed interval $z'=-1$. Since any finite $z$ value is mapped to $z'>z'_{\rm min}$ there is a sufficiently large gap to the $z'$ values associated with empty pixels. When generating new images by passing Gaussian noise through the denoising process as described above, the diffusion model does not need to generate pixels with exactly $z=0$ but, instead, it is sufficient to generate a narrow peak around $z'\sim -1$. We can then apply a lower cut at $z'_{\rm min}$ and consider pixels with smaller $z'$ values as empty. This allows us to generate sparse images of collider events. The upper gap associated with $z'_{\rm max}$ is introduced to avoid distortions of the distribution generated by the diffusion model near the upper endpoint. Any values produced by the diffusion model that are outside of the $z'\in [-1,1]$ range are clipped, which would lead to artifacts near the endpoint without including the upper gap. Note that we do not enforce a hard upper cutoff at $z'_{\rm max}$. This requires the model to learn momentum conservation, see Eq.~(\ref{eq:sumz}) above, which we quantify numerically below.

\begin{figure*}[t]
    \centering       \includegraphics[width=\textwidth]{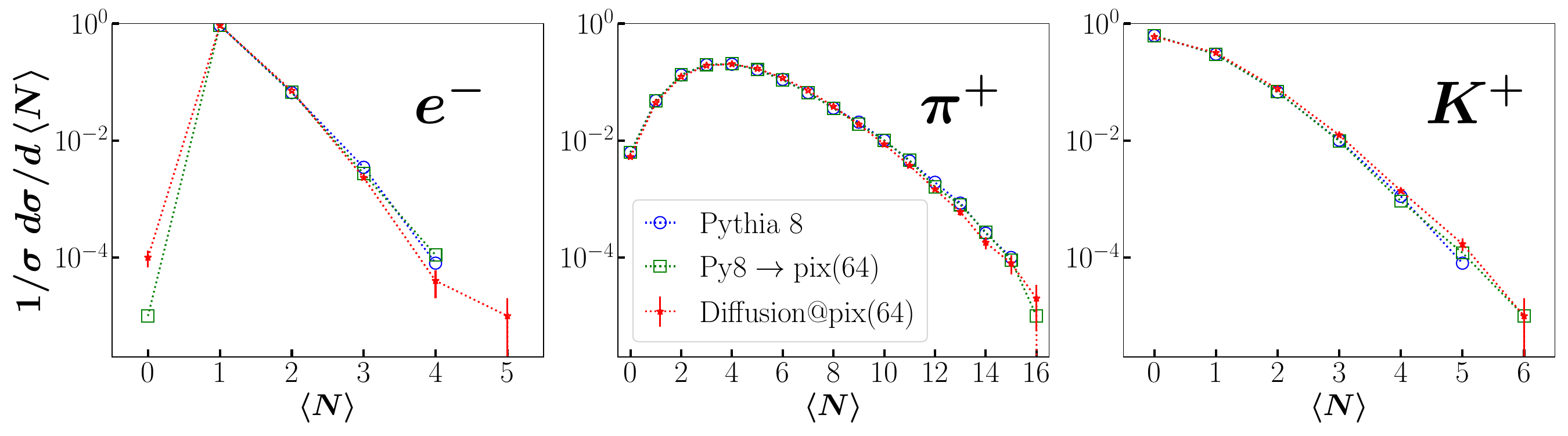}
\caption{Comparison of the (normalized) particle multiplicities $\langle N_i\rangle$ for electrons $e^-$, pions $\pi^+$, and kaons $K^+$ (from left to right). For the diffusion model results, we include statistical uncertainties.~\label{fig:mult}}
\end{figure*} 

\begin{figure}[b]
    \centering   \includegraphics[width=.47\textwidth]{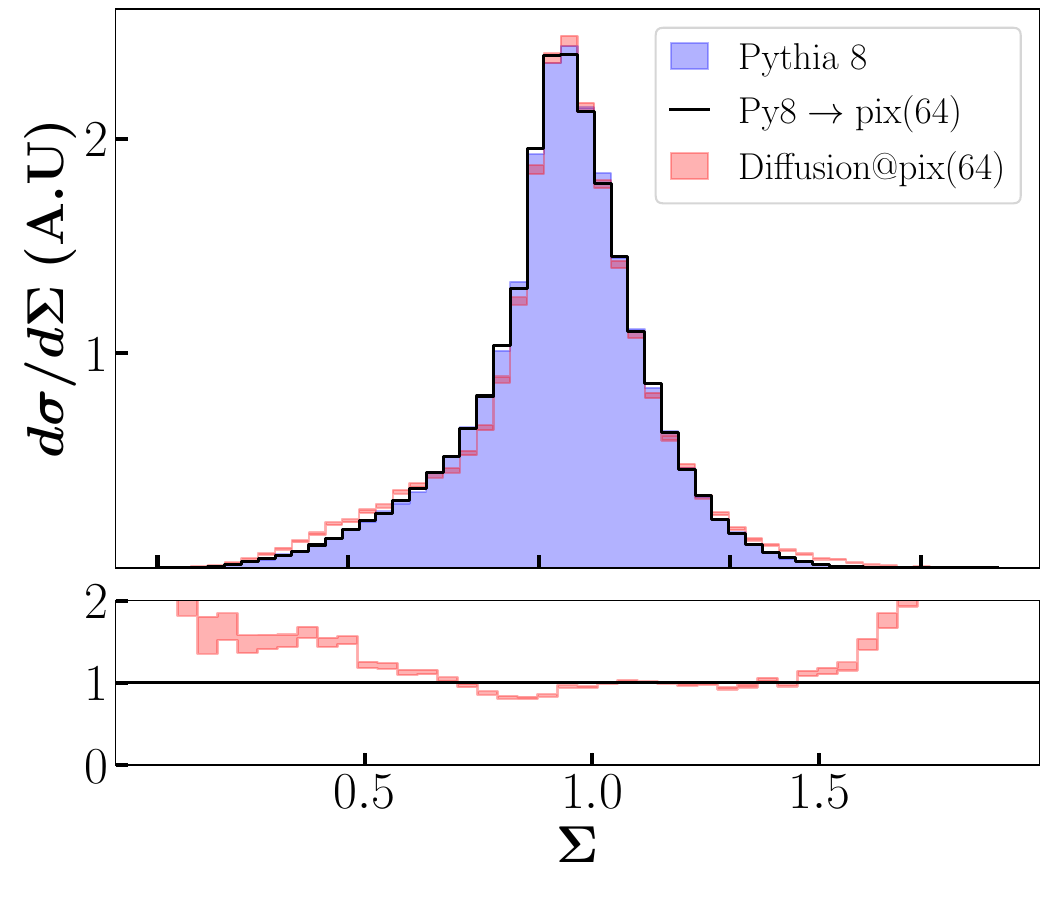}
    \caption{Comparison of \textsc{Pythia8} and diffusion model results for the sum over the rescaled momentum fractions in the entire event, see Eqs.~(\ref{eq:sumz}), (\ref{eq:zmassless}).~\label{fig:sumz}}
\end{figure}

We note that one can likely choose $E(z)=0$ and/or $S(z)=0$ for simulating proton-proton or heavy-ion collisions since in this case all particle spectra peak in the small-$z$ region. We leave the exploration of this for future work. In addition, we note that an alternative approach to simulating sparse collider data is the use of point clouds as employed in Refs.~\cite{Leigh:2023toe,Butter:2023fov,Acosta:2023zik,Leigh:2023zle,Buhmann:2023pmh} instead of the image-based data representation that we use here. 

\begin{figure*}[t]
    \centering     \includegraphics[width=\textwidth]{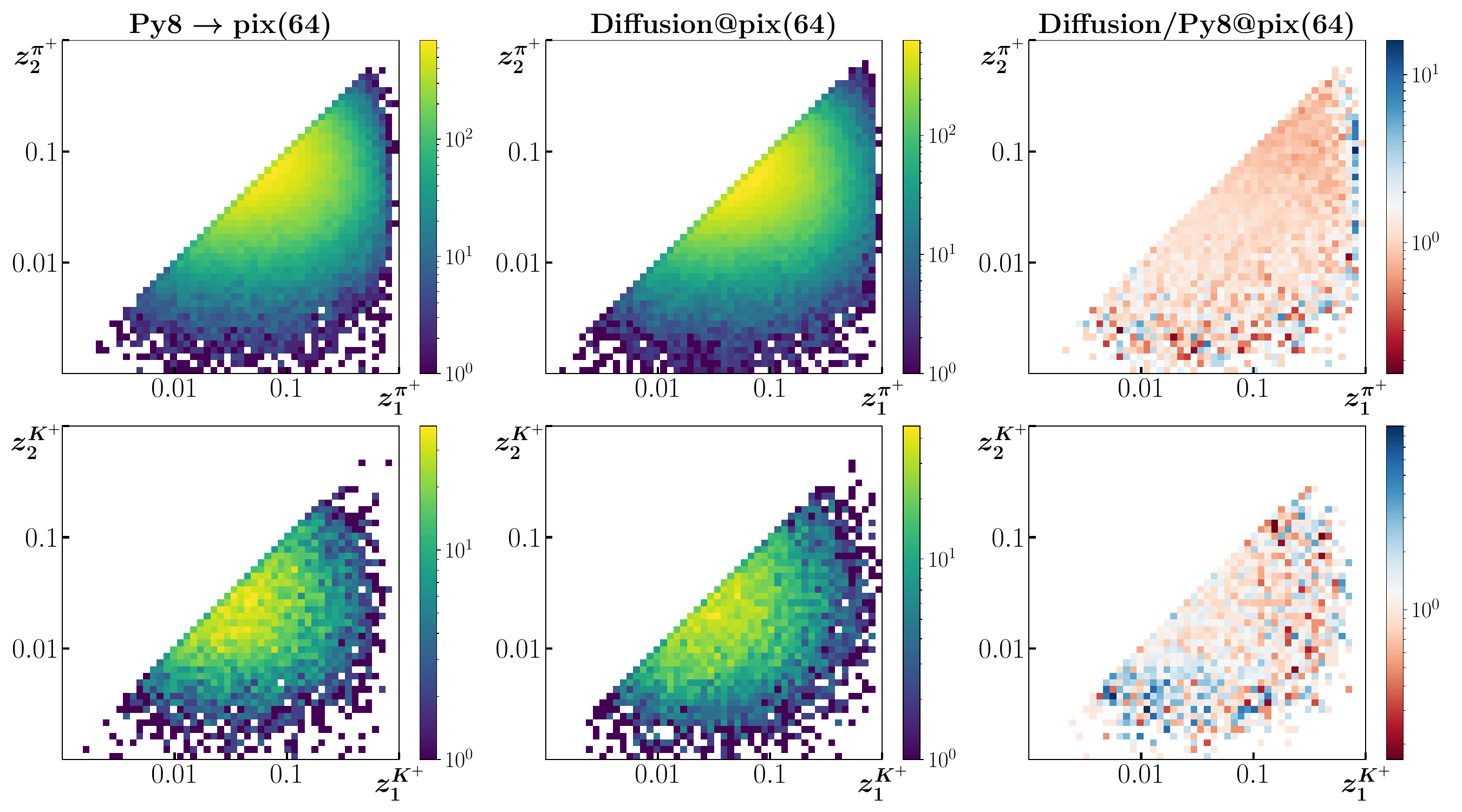}
\caption{Di-hadron correlations comparing \textsc{Pythia8} (left), the diffusion model (middle) and ratio to \textsc{Pythia8} (right): Leading and subleading pions $\pi^+$ (upper row), and kaons $K^+$ (lower row).
~\label{fig:dihadron}}
\end{figure*}

\section{Implementation and training~\label{sec:implementation}}

In this section, we are going to present more details about the implementation of the diffusion model and the training procedure. As a starting point for our work, we use the implementation of the diffusion model presented in Ref.~\cite{DBLP:journals/corr/abs-2102-09672}. In the following, we discuss aspects of the training data, the evaluation of the loss function and the parametrization of the inverse diffusion process.
\begin{itemize}
    \item The size of the training data set is $10^{6}$ images of DIS events as described in section~\ref{sec:data} above. We choose a batch size of 8.
    
    \item For the noising process $q(x_t|x_{t-1})$, we use the cosine variance schedule introduced in Ref.~\cite{DBLP:journals/corr/abs-2102-09672} with 500 diffusion steps. We find that the number of diffusion steps is sufficient for the $64\times 64$ images used in this work. The cosine variance schedule adds noise relatively slowly and is well-suited for the relatively low-resolution images considered here. For the denoising process $p_\theta(x_{t-1}|x_t)$, we use a diagonal covariance matrix $\Sigma_\theta=\sigma_t^2\mathbf{I}$, where $\sigma_t^2$ are time-dependent trainable parameters.
    
    \item We use the hybrid loss function $L_{\rm hybrid}$ given in Eq.~(\ref{eq:hybrid}) with $\lambda=0.001$ following Ref.~\cite{DBLP:journals/corr/abs-2102-09672}. Since the gradient of $L_{\rm VLB}$ in Eq.~(\ref{eq:VLB}) can be very noisy, we use importance sampling instead of uniform sampling of the this part of the objective function as proposed in Ref.~\cite{DBLP:journals/corr/abs-2102-09672}. We also explored the use of the simplified loss function proposed in Ref.~\cite{DBLP:journals/corr/abs-2006-11239} and $\lambda=0$, which corresponds to the $L_{\rm VLB}$, which generally underperformed compared to the hybrid loss function for the purposes of this work. 
    
    \item To parametrize the denoising process $p_\theta(x_{t-1},x_t)$, we use a 3-layer U-Net~\cite{DBLP:journals/corr/RonnebergerFB15} with circular or periodic convolutions. Note that this is only relevant for the azimuthal coordinate $\phi$. We choose a kernel size of 3 with stride 1 and padding 1. Multi-head attention layers~\cite{DBLP:journals/corr/VaswaniSPUJGKP17} and down/up sampling blocks are included to obtain a U-shaped neural network. It takes ${\cal O}(\text{hours})$ to generate $10^6$ samples using a single NVIDIA A100 GPU. Possible speedups can be achieved using for example the methods developed in Refs.~\cite{Buhmann:2023zgc,Mikuni:2023dvk,wu2023fast}.
    
    \item We use the AdamW optimizer~\cite{DBLP:journals/corr/abs-1711-05101} with a learning rate of $10^{-4}$. The logarithm of the loss is shown in Fig.~\ref{fig:loss} as a function of the number of samples that the diffusion model is trained on. We observe a steep decrease of the loss at the beginning. Even though the curve flattens out later during the training process, we still observe a significant improvement of the sample quality. Despite the importance sampling mentioned above, the loss turns out to be relatively noisy.
\end{itemize}

\section{Numerical results and benchmarks~\label{sec:numerics}} 

In this section, we will assess the quality of the trained diffusion model in simulating fully exclusive events (three particle species) in electron-proton collisions. We stress that our present work is limited in exploring the full extent of uncertainty quantification stemming from model uncertainties, limited training, and other factors. The results presented here should be viewed as an exploratory study, and we will focus on describing the qualitative overall agreement between the reconstructed synthetic phase space distributions and the training data. Dedicated studies of aleatoric and epistemic uncertainties are beyond the scope of our current work and will be addressed in the future.

We start with the inclusive momentum, rapidity, and azimuthal angle distributions for electrons $e^-$, pions $\pi^+$, kaons $K^+$. The comparison between the diffusion model results and \textsc{Pythia8} with and without pixelation is shown in Fig.~\ref{fig:momentum}. 
The systematic effects due to the pixelation are relatively small for the one-dimensional projections of the phase space. However, as we will find later on, they can become significant for other observables considered in this work.
Analogous to Fig.~\ref{fig:pixels}, the electron $z$-distribution peaks at $z\to 1$, whereas the pion and kaon distributions peak at small-$z$ values, see the first column of Fig.~\ref{fig:momentum}. 
We find that our mapping in Eq.~(\ref{eq:rescale}) allows us to train the diffusion model such that the $z$-spectra match relatively well. This is particularly noteworthy given the fact that the distributions of the three particle species differ substantially. However, achieving an agreement within $1\sigma$ of the estimated statistical uncertainties was not attainable. The model appears to have difficulties in reproducing the distributions near the edges of the phase while performing better in regions where the distributions are relatively flat. That being said, without the use of the mapping in Eq.~(\ref{eq:rescale}), we find that it is impossible to train the diffusion model to agree with the $z$ spectra, with deviations as large as several orders of magnitude in several regions of phase space. We conclude that a suitable mapping of the $z$-values is essential for training the diffusion model for electron-proton collisions.
The rapidity distribution for electrons shows qualitatively different features compared to pions and kaons, see the middle column of Fig.~\ref{fig:momentum}. 
The difference is again due to the unique role of the leading electron in electron-proton scattering events. The steep drop of the rapidity distributions for electrons near $\eta\sim -3$ is due to the imposed cut on the photon virtuality $Q^2$, see also Fig.~\ref{fig:ep_events}. Our model achieves a reasonable description of all three rapidity distributions with discrepancies that again tend to grow near the edges of phase space. We note that if an additional transverse momentum cut for the electron is included, the rapidity distributions for pions and electrons would be more similar. Lastly, the right column of Fig.~\ref{fig:momentum} shows the azimuthal distributions where the events have been rotated such that $\phi=0$ corresponds to the direction opposite to the leading electron, which itself is not included in these histograms. Overall, we observe that the three bell-shaped curves of the azimuthal angular correlations are approximately reproduced by the diffusion model. We observe small differences in the kaon distributions in the tails of the distribution. This is likely due to the relatively low multiplicity of kaons. Although we did not achieve full agreement within the statistical uncertainties, being able to approximately reproduce the distributions with the help of the remapping of the $z$-values encourages further explorations in this direction. Extending this work to include the estimation of model uncertainties is necessary to assess whether the observed disagreement stems from biases due to finite statistics or from a potential lack of model expressivity in our implementation.

Next, we consider particle multiplicity distributions $\langle N_i\rangle$. The comparison between \textsc{Pythia8} with and without pixelation and the diffusion model results are shown in Fig.~\ref{fig:mult}. The electron multiplicity peaks at $\langle N_e\rangle= 1$, which corresponds to the scattered leading electron, and falls off steeply as the multiplicity increases. The case of $\langle N_e\rangle= 0$ corresponds to an unphysical distortion induced by the pixelation algorithm since we are only considering neutral current DIS Pythia samples.
The pion multiplicity distribution peaks at intermediate values $\langle N_{\pi^+}\rangle\sim 4$, and exhibits a long tail extending up to $\sim 20$ pions per event. The kaon multiplicity distribution also declines rapidly toward larger values, with many events having no kaons. 

In contrast to the situation encountered for the momentum distributions, the multiplicity distributions are rather well reproduced and within 1$\sigma$ of the statistical uncertainties across most of the available phase space for the three particles. Only in the electron case, we find that the diffusion model generates non-vanishing $\langle N_e\rangle= 5$ and it systematically deviates in the unphysical case of $\langle N_e\rangle= 0$. However, these events are very rare, occurring at a rate of $10^{-4}$ relative to the peak value.

Next, we consider the sum over the rescaled momentum fractions, see Eq.~(\ref{eq:sumz}) above. This provides an important test of the global characteristics of the events generated by the diffusion model. The diffusion model result compared to \textsc{Pythia8} is shown in Fig.~\ref{fig:sumz}. 
%
%Overall, we observe good agreement for the event-wide momentum sum rule in Eq.~(\ref{eq:sum}).
%
The distribution peaks near $\sum_i z_i\sim 1$ and falls off steeply toward the endpoints, where the upper limit results from momentum conservation, see Eq.~(\ref{eq:sumz}) above. It is worth noting that while we simulate pions $\pi^+$ and kaons $K^+$, which correspond to some of the most frequently produced particles in the events, the distributions here are significantly shifted to the left compared to the upper boundary due to the omission of other particle species (e.g. $\pi^-,\pi^0$) in our current implementation.
Our diffusion model is able to approximately reproduce this distribution. While the peak around $\Sigma\sim 1$ is well reproduced, the deviations grow toward the edges of phase space. We find that the model does not generate unphysical events with $\Sigma >2$. We conclude that the model is able to approximately learn momentum conservation, i.e. without it being imposed directly as an additional constraint.

\begin{figure*}[t]
    \centering   
    \includegraphics[width=\textwidth]{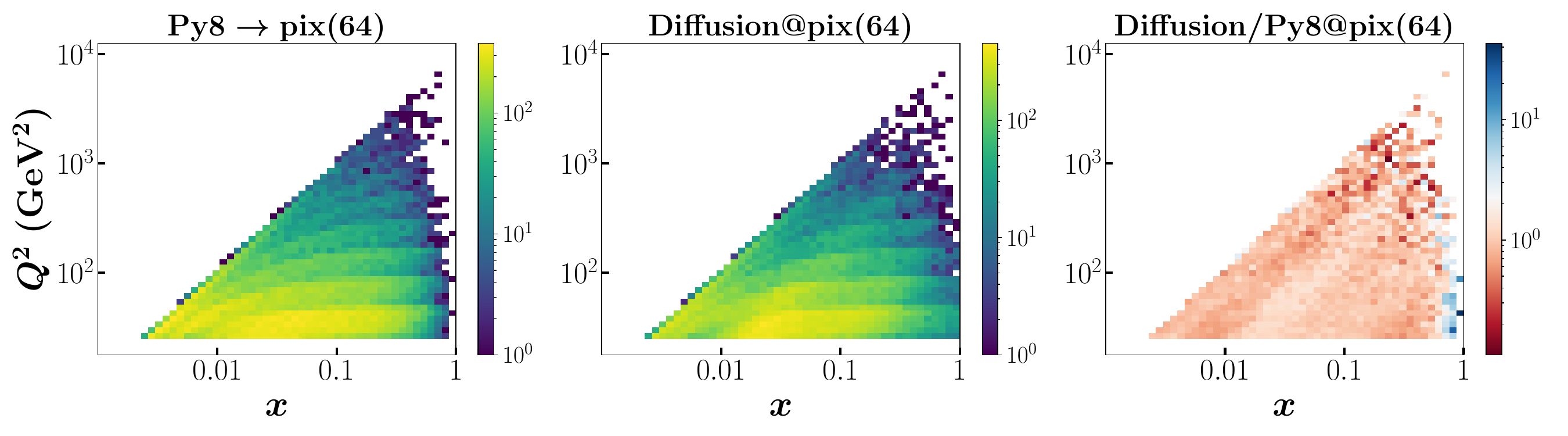}
    \caption{Correlations of the DIS variables Bjorken  $x$ and the photon virtuality $Q^2$ comparing \textsc{Pythia8} events (left) to results from the diffusion model (middle) and ratio to \textsc{Pythia8} (right).}
    \label{fig:xQ2}
\end{figure*}

Next, we consider a set of multi-hadron correlations, which serve as an important benchmark for evaluating the ability of the diffusion model to capture correlated features beyond the one-dimensional projections shown earlier. The two-dimensional histograms of the momentum fractions of leading versus subleading pions (upper row) and kaons (lower row) are shown in Fig.~\ref{fig:dihadron}. The diffusion model clearly learns the correlations between the leading and subleading particles that vary within two orders of magnitude in phase space regions with large statistics and loses its ability to reproduce the correlations toward the edges of phase space. As mentioned before, quantifying the model uncertainty is required to better assess the degree of agreement. However, we conclude that our results in Fig.\ref{fig:dihadron} encourage future explorations in this direction.

We are now going to evaluate the diffusion model's performance in describing inclusive DIS reactions characterized by the virtuality of the exchanged photon $Q^2=-q^2=-(l-l')^2$, and the Bjorken scaling variable, defined as $x=Q^2/2p\cdot q\leq 1$. Here $l,l'$ denote the incoming and outgoing four-momenta of the scattered electron, respectively, and $p$ is the momentum of the incoming proton. Assessing the diffusion model's ability to reproduce the distribution of these variables is a stringent test that allows us to gauge its ability to capture the correlations between the outgoing electron phase space and the initial state momenta. In Fig.~\ref{fig:xQ2} we display the reconstructed 2D density in the $x,Q^2$ space. Overall, we find good agreement between the diffusion model results and \textsc{Pythia8} with pixelation effects with some noticeable differences only in the $x\to 1$ region. The presence of stripes in the 2D density plots can be attributed to pixelation effects, which can be mitigated by increasing the number of pixels used to represent the events, analogous to the momentum distributions shown in Fig.~\ref{fig:pixels}. We defer improvements to future work due to limitations of computational resources. Nevertheless, within the described limitations, we conclude that the diffusion model is able to approximately reproduce the correlations of the DIS phase space.

As discussed above, we use the variable $z$ as the pixel intensity rather than the transverse momentum $p_{T}$. Here, we discuss why the pixelation of $p_{T}$ induces larger systematic uncertainties compared to $z$ when reconstructing the DIS kinematic variables $x$ and $Q^2$. The photon virtuality can be expressed as $Q^2=2l_0l'_{T}e^{-y_{l'}}$, where $l_0$ is the incident lepton energy, and $l'_{T}$ and $y_{l'}$ are the outgoing lepton transverse momentum and rapidity in the lab frame, respectively. Focusing on the uncertainty induced by the rapidity pixelation, we find that $\delta Q^2\sim Q^2 \delta y_{l'}$ for fixed values of $l'_{T}$. This implies that uncertainties on $Q^2$ from the rapidity pixelation are amplified by a factor of $Q^2$, which is typically required to be large for phenomenological applications. Instead, when using the variable $z_i$, we find the that the photon virtuality is given by $Q^2=l_0 \sqrt{s} z (e^{-y_{l'}}/\cosh y_{l'})$, where $\sqrt{s}$ is the CM energy. In this case, the uncertainty induced by the rapidity pixelation for fixed values of $z$ is $\delta Q^2=Q^2(e^{y_{l'}}/\cosh{y_{l'}})\delta y_{l'}$, i.e. there is an additional factor $K=(e^{y_{l'}}/\cosh{y_{l'}})$ relative to the $p_{T}$ pixelation case. In the CM frame of the electron-proton reaction, with the incoming beam of electrons moving along the $z$-axis, the rapidity of the outgoing electrons is mostly negative, and typical values for $K$ are around 0.01, which significantly suppresses systematic errors.

\begin{figure*}[t]
    \centering   
    \includegraphics[width=\textwidth]{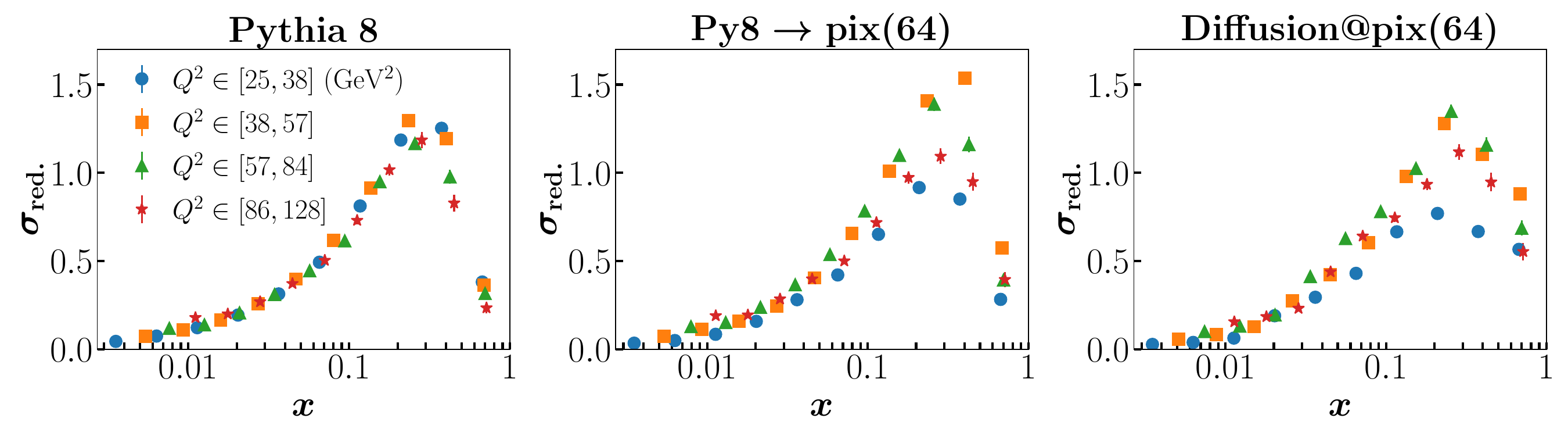}
    \caption{Photon virtuality $Q^2$ scaling of the DIS reduced cross sections as a function of Bjorken $x$. The three panels show the \textsc{Pythia8} results without and with pixelation and the diffusion model results (from left to right).~\label{fig:sigma_red}}
\end{figure*}

One of the salient features of the DIS process is the presence of scaling in the variable $Q^2$ which is interpreted as the evidence of point-like constituents inside the proton. The so-called DIS neutral current (NC) reduced cross sections is defined as  
\begin{align}
    \sigma_{\rm red.~NC}^{ep\to e'+X} &= \frac{{\rm d}\sigma^{ep\to e'+X}_{\rm NC} }{{\rm d}x{\rm d}Q^2}\frac{Q^4 x}{2\pi \alpha^2 Y_+ }\notag\\ 
    &= F_2(x,Q^2)- \frac{y^2}{Y_+}F_L(x,Q^2) + \frac{Y_-}{Y_+}xF_3(x,Q^2) \;.
\end{align}
Here we defined $Y_{\pm} =1\pm(1-y)^2$, with the inelasticity $y=Q^2/(s-M^2)/x$. $M$ is the proton mass, and $\alpha$ corresponds to the electromagnetic fine-structure constant. The structure functions $F_{2,L,3}$ are independent of $Q^2$ up to logarithmic corrections that can be predicted within perturbative QCD, provided that $Q^2$ is sufficiently large relative to any other hadronic scale. In this regime the $F_2$ structure function is the dominant contribution to the reduced cross section, hence it is approximately invariant under changes in $Q^2$. In Fig.~\ref{fig:sigma_red} we compare the reconstructed cross section $\sigma_{\rm red.~NC}^{ep\to e'+X}$ from diffusion model generated events and the \textsc{Pythia8} samples. 
The reconstructued reduced cross sections from \textsc{Pythia8} shows the expected scaling behaviour. In contrast after pixelation,  distortions on the scaling behavior are induced mostly in the large-$x$ valence region. We stress that this can be mitigated by enlarging the number of pixels. Taking into account these systematic scaling violations induced by the pixelation, the reconstructed reduced cross sections from the diffusion model are qualitatively in agreement with the pixelized version of \textsc{Pythia8}, albeit the diffusion model exhibits deviations at high values of $x$ compared to \textsc{Pythia8}. These deviations may be associated with the epistemic uncertainties of the diffusion model. We would like to highlight that achieving a faithful representation of DIS events serves as the starting point for studying cross sections such as semi-inclusive DIS that are differential in up to approximately 10 variables, which will play a critical role at the future EIC. We note that further improvements of our results may be achieved by using diffusion models based on point clouds~\cite{Acosta:2023zik} and adapting methods developed in Ref.~\cite{Nachman:2023clf} in the context of emulating hard-scattering events.

\section{Conclusions~\label{sec:conclusions}} 

In this work, we presented simulations of electron-proton scattering events using diffusion models. Our results are relevant for simulations at CEBAF, the future Electron-Ion Collider and LHeC/FCC-eh. The diffusion model is based on a noising schedule that sequentially turns the images from the training data set into Gaussian noise. The stochastic reverse process is learned by a U-Net architecture based on convolutional layers with small filters. We trained the diffusion model on \textsc{Pythia8} simulations of electron-proton scattering events at EIC energies and observed that it can generate high-quality sparse samples of collider events. We achieved good agreement near the kinematic endpoints by rescaling the particle momenta with a mixed exponential-logarithmic function, which accounts for the unique role that the scattered electron plays in electron-proton collisions. We employed an image-based representation of the training data and explored associated pixelation effects. As a first step, we limited ourselves in this work to electrons, pions, and kaons, which are represented by different ``color'' channels. Overall, we found good agreement for various observables and their correlations as well as event-wide constraints such as momentum conservation. We foresee various applications of our work in the context of generative modeling for collider physics including event-level analysis of hadron structure, data storage, studies of hadronization, exclusive processes like Deeply Virtual Compton Scattering (DVCS), and searches of new physics, which will be addressed in future work.

\begin{acknowledgements}
We would like to thank Jack Araz, Prerit Jaiswal, Vinicius Mikuni, James Mulligan,  Mateusz Ploskon, and Fernando Torales-Acosta for helpful discussions. PD, JWQ, FR, and NS are supported by the U.S. Department of Energy, Office of Science, Contract No.~DE-AC05-06OR23177, under which Jefferson Science Associates, LLC operates Jefferson Lab. FR is supported in part by the DOE, Office of Science, Office of Nuclear Physics, Early Career Program under contract No~DE-SC0024358. N.S. is supported by the DOE, Office of Science, Office of Nuclear Physics in the Early Career Program. 
\end{acknowledgements}

\bibliographystyle{utphys}
\bibliography{main.bib}

\end{document}